*Paper presented at EUCAS'01*

# Superconducting properties of the powder-in-tube Cu-Mg-B and Ag-Mg-B wires

B.A.Glowacki [a,b], M.Majoros [b] M.E.Vickers [a] and B.Zeimetz [a]
[a] Department of Materials Science and Metallurgy, University of Cambridge, Pembroke Street, Cambridge CB2 3QZ, UK,
[b] Interdisciplinary Research Centre in Superconductivity, University of Cambridge, Madingley Road, Cambridge CB3 OHE, UK

## Abstract

The new class of Mg-B superconducting conductors was prepared using the standard *in-situ* powder-in-tube, PIT, method where Mg+2B mixture was used to fill the copper and silver tubes. Study of the intergranular current, grain connectivity and superconducting phases in wires were conducted by AC susceptibility measurements and direct four point transport measurements. Using SQUID magnetometer, magnetisation versus magnetic field (M-H) curves of the round wires after reactive diffusion were measured at temperatures 5K and magnetic field up to 5T to define the $J_{cmag}$. The direct current density measurements of the in-situ Cu-$MgB_2$ performed at 4.2K gave the value of $10^9 Am^{-2}$.

*B.A.Glowacki , Department of Materials Science and Metallurgy, University of Cambridge, Pembroke Street, Cambridge CB2 3QZ, UK, fax 00441223334567 e-mail bag10@cam.ac.uk*

## 1. Introduction

Commercial exploitation of the recently discovered $MgB_2$ superconductors will be severely limited unless mechanically robust, high critical current density, composite conductors can be fabricated with uniform properties over long lengths.

There are two different conductor preparation methods: an *in situ* technique where an Mg+2B or $MgB_2$+(Mg+2B) mixture can be used as a central core of the powder-in-tube, PIT, conductor or an *ex-situ* technique where fully reacted $MgB_2$ powder which may be doped or chemically modified can be used to fill the metal tube [1-3].

## 2. Experimental and results

*2.1 In-situ Cu/MgB$_2$ wires*

All samples were prepared by a mixture of Mg and B powders in stoichiometric amounts and made by PIT method in Cu tubes. Outside diameters of the wires ranged from 1 to 2mm and the cross sectional $MgB_2$ to metal ratio was ~0.3. Wire #1 was sintered at 620$^{o}$C for 48 hours in vacuum, wire #2 was sintered at 700$^{o}$C for 1 hour in argon atmosphere (ramp rate of the furnace 300$^{o}$C/h) and wire #3 was sintered at 800$^{o}$C for 1 hour in argon atmosphere (ramp rate of the furnace 300$^{o}$C/h). For ac susceptibility measurements pieces of the samples about 3 mm in length were used. Magnetic field was applied parallel to the axis of the wires. Direct four point current measurements were conducted at 4.2 K in self field.

Wires manufactured by *in-situ* technique, diffusing Mg to B particles experienced ~25.5% decrease in density from the initial density value after cold deformation, due to the phase transformation from Mg +2($\beta$−B) =>$MgB_2$ (all hexagonal structures).

The internal susceptibility $\chi_{int}$ is shown in Figure 1 for all wires. The susceptibilities were normalised so as to have $\chi_{int}$ =0 at 100 K. Wire #1 and wire #2 have a sharp transition with onset $T_c$=39K. In all cases, some anomalous decrease of the susceptibility at T~50 K can be seen. Wire #3 has a broad transition with onset $T_c$~50 K and downset $T_c$~25 K. There is some frequency dependence of the susceptibility due to skin effect in copper. At 50 K the resistivity of Cu is $\rho$=4x10$^{-10}$ $\Omega$m [4] and the skin depth $\delta=(\rho/\mu_o\pi f)^{0.5}$ at frequencies $f$=333.3 Hz starts to be comparable with wall thickness of the copper tube of our samples, which is about 0.25 mm. To analyse possible screening effects we grounded the inner core of the wire #2 and wire #3 into a powder and measured ac susceptibilities again. The results still have an anomalous decrease of the susceptibility at $T$~50K [3]. DC SQUID measurements performed on wire #3 in form of a powder confirmed the anomalous magnetic moment decrease at *T*~50K [3].

Our measurements using a microanalyser with wavelength dispersive spectrometers, MAWDS, conducted on the wire #2 showed that as a result of Cu-Mg-B interaction multiphase conductors have been obtained and there is no interdiffusion between Cu and B, see Fig. 2. X-ray powder pattern conducted on the powdered cores of the conductors shows a mixture of $MgB_2$ and other phases, namely $MgCu_2$, MgO, $MgB_4$ and possibly also higher borides. With increasing temperature mor Mg diffused into the copper matrix.

Calculations were performed to investigate the effect, on the X-ray powder data, of Cu substitution into the $MgB_2$ structure. Several models were considered and the two main ones discussed here: 1) 20%Cu substituted randomly onto Mg sites and 2) 40% Cu on every second Mg layer thus increasing c from 3.521 to 7.04Å. Comparing the data calculated from model 1 to that of $MgB_2$ there was a small variation in the relative intensity of some peaks, the greatest being about 10% for (010) at ~33.5-2θ. Model 2 gives two additional peaks: (001) at 12.56 and (011) at 35.94-2θ. The lower peak was never observed in our samples thus this model must be excluded. However, a peak at about 35.9-2θ was observed and was the reason for considering a doubled cell. This can be indexed as the (022) for $Cu_2Mg$ as calculated from recent data [5] but not reported in the old PDF data no 1-1226. Thus although model 2 should be excluded, the possibility of low levels of Cu substituted into the $MgB_2$ structure cannot be ruled out.

Rietveld refinement [6] was carried out on several samples including one where attempts were made to remove some of the metal giving a sample with a higher concentration of black powder. The experimental data were fitted with three crystalline phases: Cu, $MgCu_2$ and $MgB_2$ or $Mg_{0.8}Cu_{0.2}B_2$ or the layered model mentioned earlier. In every case the main difficulty was fitting the $MgCu_2$ phase which could be somewhat disordered or perhaps there is a problem with absorption. The results for $MgB_2$, $Mg_{0.8}Cu_{0.2}B_2$ and the layered model were: respectively profile ($R_p$) 6.05, 6.03 and 6.02 and goodness of fit 101.7,101.7 and 101.0. Within experimental error these refinements are the same. Thus as stated earlier, the layered model should be excluded because no peak was observed at about 12.5-2θ but these data, with considerable overlap of the $MgB_2$ and $MgCu_2$ peaks, cannot differentiate between $MgB_2$ and $Mg_{0.8}Cu_{0.2}B_2$.

Using a SQUID magnetometer, magnetisation versus magnetic field (*M-H*) curves of the round *in situ* Cu clad $MgB_2$ wires after sintering in the range of temperatures and time were measured at temperature of 5 K and magnetic fields up to 5 T to define the $J_{cmag}$, see Fig 3. The transport data for the Cu claded wire is incorporated in Fig 3. Use of copper also minimises the overall cost of the tape or round wire, because expensive non magnetic barrier materials ( such as Nb or Ta [7]) surrounding the $MgB_2$ phase, giving further room to improve the conductor stabilisation. The reproducibility of the measured transport currents was good and not dependent on the wire diameter.

*2.2 Ag-$MgB_2$ conductors*

In the case of an *in-situ* wires made in silver tubes some anomalous decrease of the susceptibility at T~50 K was also observed. It appeared that *in-situ* wires have a broad transition and almost zero transport current [8]. AC susceptibility measurements of the *ex-situ* wires additionally sintered also presented anomalous decrease of the susceptibility at T~50 K which disappeared with increasing sintering temperature, Fig.4. Sintering at lower temperatures improved intergrain connectivity of the as drawn *ex-situ* wire. However further increase of the temperature caused degradation of the superconductor by possible extensive reactive diffusion of Mg to silver matrix. The transport critical current for the *ex-situ* silver claded wires was considerably lower than that for copper PIT conductors.

## 3. Conclusions

Cu/(Mg-2B) *in-situ* conductors present a low cost fully stabilised potential solution for magnet applications. Further systematic investigation of copper and silver possible substitution in Mg borides using STEM with high resolution X-ray diffraction and at Daresbury synchrotrons (UK) are going to be performed in the near future to understand the origin of 50K anomalies.

## Acknowledgement

M.Majoros acknowledges the AFRL/PRPS Wright-Patterson Air Force Base, Ohio, USA for the financial support.

## References

[1] B.A. Glowacki, M. Majoros, M.Vickers, J.E. Evetts, Y. Shi and I. McDougall, Supercond. Sci. Technol. 14 (2001) 193.

[2] W. Goldacker, S.I. Schlachter, S. Zimmer, H. Reiner, cond-mat/0106226, also submitted to

SST.
[3] B.A.Glowacki M.Majoros,In: Studies of superconductors (Advances in Research and Applications) $MgB_2$ ed. A.Narlikar, Nova Science Publishers, Inc., Huntington, New York, 2001, v38. p.361.
[4] G.K. White, Experimental techniques in low-temperature physics, Oxford, Clarendon Press (1959).
[5]T. Ohba, Y. Kitano and Y. Komura Acta. Cryst. C 40 (1984) 1.
[6] Philips X'pert Plus software implenting a modified version of Wiles and Young code J. Appl. Cryst. 14 (1981) 149.
[7] W. Goldacker, S.I. Schlachter, S. Zimmer, H. Reiner, cond-mat/0106226, Supercond. Sci. Technol. (in press).
[8] M. Majoros and B.A. Glowacki, submitted to Supercond. Sci. Technol.

## Figure captions

Figure 1 The AC susceptibility of in situ Cu/(Mg-2B) wire sintered at different temperatures under protective argon atmosphere, measured at frequency 333.3 Hz in applied magnetic field 1Oe (rms). The arrow indicates position of the change in susceptibility slope at ~50K common for all samples.

Figure 2 Scanning Electron Micrograph and MAWDS surface analysis of Cu-$MgB_2$ interface region in a in situ Cu/(Mg-2B) conductor sintered at 700$^{o}$C for 1h.

Figure 3 The transport and magnetic critical current density, $J_{cmag}$, defined from magnetisation, (M-H) curves of the round *in situ* Mg-2B Cu/(Mg-2B) wires after reactive sintering formation of $MgB_2$; samples were measured at 5K, using a SQUID magnetometer. The measured transport current density $J_{ctrans}$ ~$10^5$Acm$^{-2}$ is represented by solid square which was calculated for the whole cross section of the B rich area superconducting core diameter, however it is an underestimated value since only ~70% of the whole cross section can be estimated to be an effective superconductor, see Fig. 2.

Figure 4 The AC susceptibility of in situ Ag-($MgB_2$) wires sintered at different temperatures under vacuum, measured at frequency 33.3 Hz in applied magnetic field 1Oe (rms). The arrow indicate position of the change in susceptibility slope at ~50K mostly pronounced for samples sintered at lower temperatures. The origin of curves has been shifted down in steps of 0.033 form a 0 position at 100K to underline the anomalies at 50K.

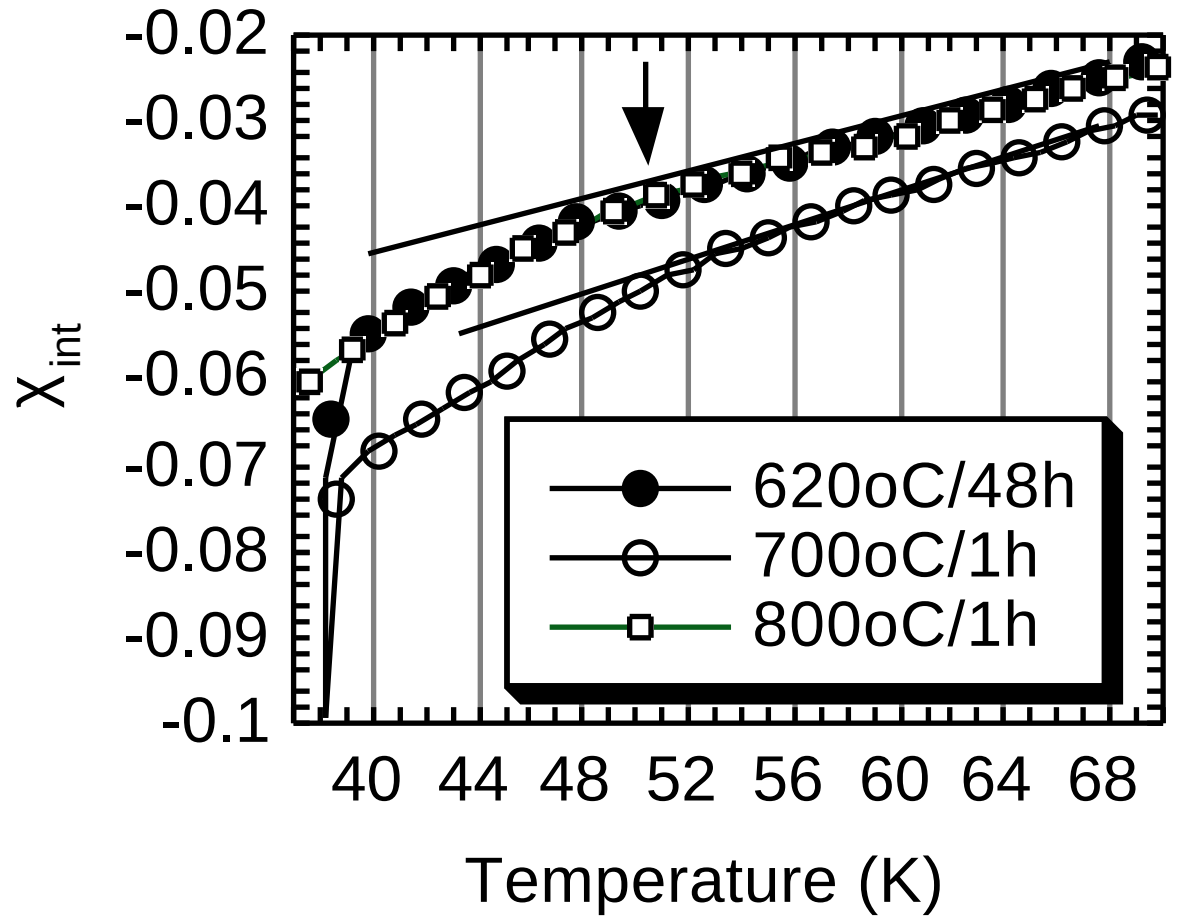

Fig.1 B.A.Glowacki et al. ‘Superconducting properties of the powder-in-tube Cu-Mg-B ....’ No. N1.2-07

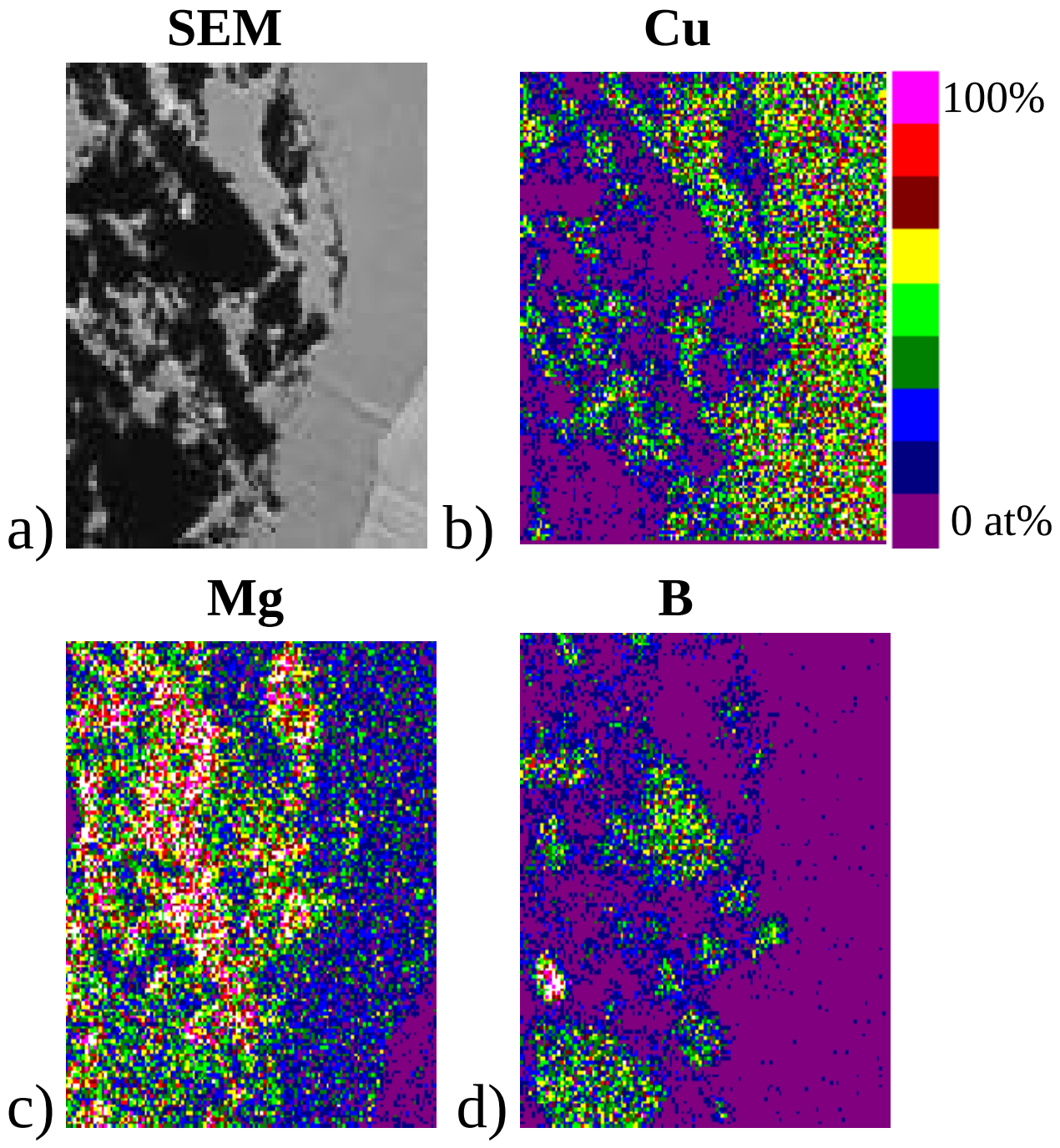

Fig.2 B.A.Glowacki et al. 'Superconducting properties of the powder-in-tube Cu-Mg-B ....' No. N1.2-07

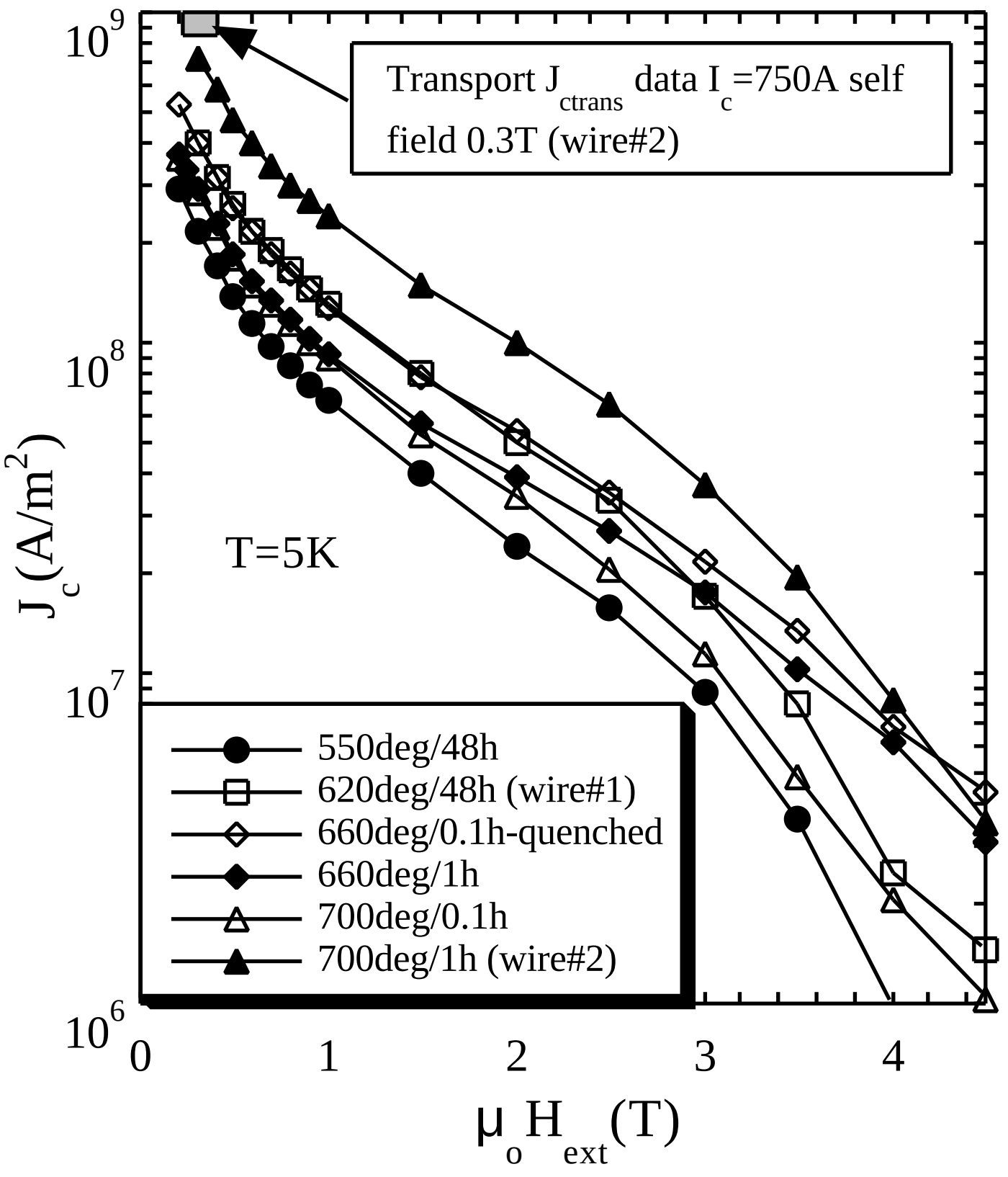

Fig.3 B.A.Glowacki et al. 'Superconducting properties of the powder-in-tube Cu-Mg-B ....' No. N1.2-07

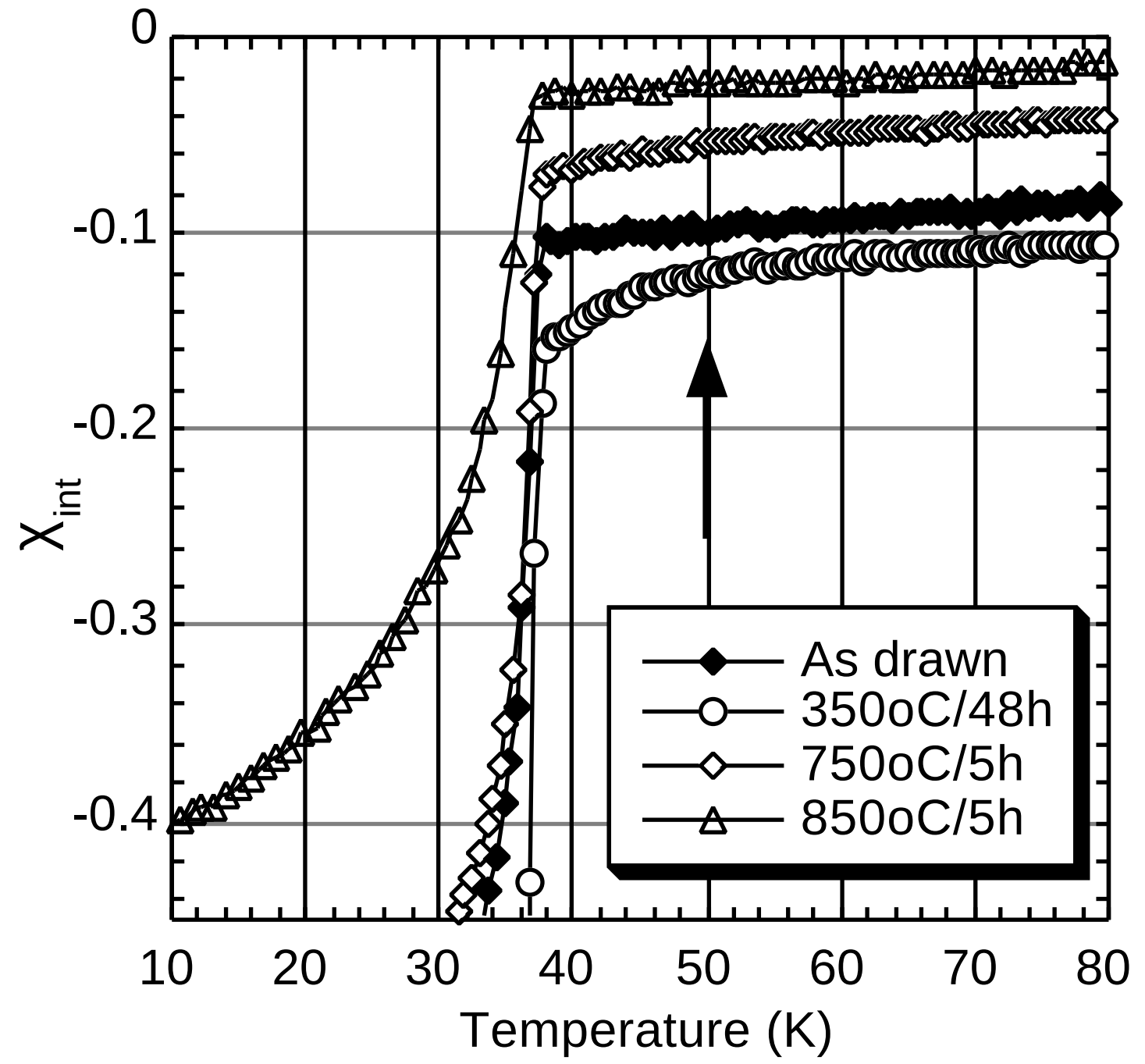

Fig.4 B.A.Glowacki et al. 'Superconducting properties of the powder-in-tube Cu-Mg-B ....' No. N1.2-07